\documentclass[prl,twocolumn,latterpaper,superscriptaddress, 10pt]{revtex4-1}  

\usepackage{graphicx}  
\usepackage{color}
\usepackage{dcolumn}   
\usepackage{bm}        
\usepackage{amssymb}   
\usepackage{amsmath,textcomp}
\usepackage{xfrac}
\usepackage[utf8]{inputenc}
\usepackage{float}
\usepackage{hyperref}
\usepackage[margin = 1in]{geometry}

\begin{document}

\title{Observation of a room-temperature oscillator's motion dominated by quantum fluctuations over a broad audio-frequency band}

\author{Jonathan Cripe}
\email{jcripe1@lsu.edu}
\affiliation{Department of Physics \& Astronomy, Louisiana State University, Baton Rouge, LA, 70803}

\author{Nancy Aggarwal}
\affiliation{LIGO - Massachusetts Institute of Technology, Cambridge, MA 02139}

\author{Robert Lanza}
\affiliation{LIGO - Massachusetts Institute of Technology, Cambridge, MA 02139}

\author{Adam Libson}
\affiliation{LIGO - Massachusetts Institute of Technology, Cambridge, MA 02139}

\author{Robinjeet Singh}
\affiliation{Department of Physics \& Astronomy, Louisiana State University, Baton Rouge, LA, 70803}

\author{Paula Heu}
\affiliation{Crystalline Mirror Solutions LLC and GmbH, Santa Barbara, CA, 93101 and 1060 Vienna, Austria}

\author{David Follman}
\affiliation{Crystalline Mirror Solutions LLC and GmbH, Santa Barbara, CA, 93101 and 1060 Vienna, Austria}

\author{Garrett D. Cole}
\affiliation{Crystalline Mirror Solutions LLC and GmbH, Santa Barbara, CA, 93101 and 1060 Vienna, Austria}
\affiliation{Vienna Center for Quantum Science and Technology (VCQ), Faculty of Physics, University of Vienna, A-1090 Vienna, Austria}

\author{Nergis Mavalvala}
\affiliation{LIGO - Massachusetts Institute of Technology, Cambridge, MA 02139}

\author{Thomas Corbitt}
\affiliation{Department of Physics \& Astronomy, Louisiana State University, Baton Rouge, LA, 70803}


	\date{\today}
\date{\today}

\begin{abstract}

We report on the broadband measurement of quantum radiation pressure noise (QRPN) in an optomechanical cavity at room temperature over a broad range of frequencies relevant to gravitational-wave detectors. We show that QRPN drives the motion of a high-reflectivity single-crystal microresonator, which serves as one mirror of a Fabry-P\'erot cavity.
In our measurements QRPN dominates over all other noise between 10 kHz and 50 kHz and scales as expected with the circulating power inside the cavity. The thermal noise of the microresonator, the largest noise source next to the QRPN, is measured and shown to agree with a structural damping model from 200 Hz to 30 kHz. By observing the effects of QRPN in the audio-band, we now have a testbed for studying techniques to mitigate back-action, such as variational readout and squeezed light injection, that could be used to improve the sensitivity of gravitational-wave detectors.

\end{abstract}

\maketitle


Quantum mechanics places a fundamental limit on the precision of continuous measurements. The Heisenberg uncertainty principle dictates that as the precision of a measurement of an observable (e.g. position) increases, back action creates increased uncertainty in the conjugate variable (momentum). In gravitational wave (GW) interferometers, the laser power is increased as much as possible to reduce the position uncertainty created by shot noise but at the expense of back-action in the form of quantum radiation pressure noise (QRPN) \cite{Caves_1980}. Once at design sensitivity, Advanced LIGO \cite{LIGO}, VIRGO \cite{VIRGO}, and KAGRA \cite{KAGRA} will be limited by QRPN at frequencies between 10 Hz and 100 Hz.  To improve the detection rate of GWs, ideas have been proposed to mitigate the QRPN  \cite{Braginsky547, Braginsky_speedmeter, Kimble, Harms, Filter, FD_SQZ, Glasgow_speedmeter}, but until now there has been no platform to experimentally test these ideas. Here we present a broadband measurement of QRPN at room temperature at frequencies relevant to GW detectors. The measured noise spectrum shows effects from the QRPN between about 2 kHz to 90 kHz, and the measured magnitude of QRPN agrees with our model. 
We now have a testbed for studying techniques to mitigate back-action, such as variational readout and squeezed light \cite{Kimble}, that could be used to improve the sensitivity of GW detectors.

Gravitational wave detectors such as Advanced LIGO continuously monitor the position of test masses using electromagnetic waves. In this case the Heisenberg uncertainty principle requires that $\Delta N \cdot \Delta \phi \geq \sfrac{1}{2}$, where $\Delta N$ is the uncertainty in the number of photons and $\Delta \phi$ is the uncertainty in the phase. The photon number uncertainty exerts back-action (QRPN) on the mirrors on reflection,  causing them to vibrate \cite{Caves_1980, Braginsky_book, Caves1981}. GW interferometers typically use as much laser power as possible in order to minimize the phase uncertainty and maximize the signal-to-noise ratio for GWs.  At sufficiently high powers, however, the QRPN becomes larger than the phase uncertainty, and it is no longer advantageous to further increase the laser power. Advanced LIGO will be limited by QRPN at low frequency when running at its full laser power.  


 Given the imperative for more sensitive GW detectors, it is important to study the effects of QRPN in a system similar to Advanced LIGO, which will be limited by QRPN across a wide range of frequencies far from the mechanical resonance frequency of the test mass suspension. Studying quantum mechanical motion is challenging, however, due to the fact that classical noise sources such as environmental vibrations and thermally-driven fluctuations \cite{Saulson} usually dominate over quantum effects. Previous observations of QRPN have observed such subtle quantum effects, even at room temperature, but these experiments have thus far been limited to high frequencies (MHz-GHz) and in a narrow band around a mechanical resonance \cite{Purdy, Teufel, Purdy_room_T, Sudhir}.

\begin{figure}
\center
\includegraphics[width=1\columnwidth]{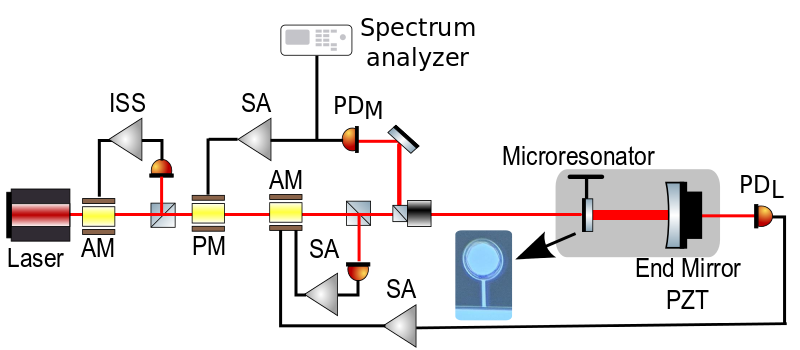}
\caption{Experimental  Setup. Light from a 1064 nm Nd:YAG laser is passed through an amplitude modulator (AM), phase modulator (PM), and second AM before being injected into the in-vacuum optomechanical cavity, which sits on a suspended optical breadboard to reduce seismic motion (shown in shaded grey).
A micrograph of the single-crystal microresonator, comprising a 70-$\mu$m diameter GaAs/AlGaAs mirror pad supported by a GaAs cantilever, is included in the diagram.
An intensity stabilization servo (ISS) and secondary servo are used to maintain a constant input power by feeding back to the AMs. The light transmitted through the optomechanical cavity is detected by $\mathrm{PD_L}$. The signal from $\mathrm{PD_L}$ is sent through a servo amplifier (SA) before being sent to the second AM to initiate the cavity lock sequence. The beam reflected by the cavity is detected by photodetector $\mathrm{PD_M}$. The signal from $\mathrm{PD_M}$ is used to lock the cavity by sending it through a separate SA feeding back to the PM. The $\mathrm{PD_L}$ locking loop is turned off after the $\mathrm{PD_m}$ locking loop is active. The signal from $\mathrm{PD_M}$ is also sent to a spectrum analyzer for further analysis. 
}
\label{Setup}
\end{figure}

In this work, we present a broadband and off resonance measurement of QRPN in the audio frequency band.
We have developed low-loss single-crystal microresonators with sufficiently minimized thermal noise at room temperature such that the quantum effects can be observed. The optomechanical system, shown in detail in Figure \ref{Setup}, is a Fabry-P\'erot cavity with a mechanical oscillator as one of the cavity mirrors. The optomechanical cavity is housed in a vacuum chamber and sits on a suspended optical breadboard to provide isolation from seismic motion. The optomechanical cavity is slightly less than 1-cm long and consists of a 
high-reflectivity single-crystal microresonator that serves as the input coupler and a macroscopic mirror with a 1-cm radius of curvature as the back reflector. The cavity is made slightly shorter than the 1-cm radius of curvature of the large mirror in order to achieve a small spot size on the microresonator while maintaining stable cavity modes.
The microresonator consists of a roughly 70-$\mu$m diameter mirror pad suspended from a single-crystal GaAs cantilever with a thickness of 220-nm, width of 8-$\mu$m, and length of 55-$\mu$m. The mirror pad is made up of 23 pairs of quarter-wave optical thickness GaAs/Al$_{ 0.92}$Ga$_{0.08}$As layers for a transmission of $T = 250$ ppm and exhibits both low optical losses and a high mechanical quality factor \cite{cole08, cole12, cole13, cole14, Singh_PRL}. The microresonator has a mass of 55 ng, a natural mechanical frequency of $\Omega_m=2\pi\times876$ Hz, and a measured mechanical quality factor $Q_m=16,000$ at room temperature. The cavity has a finesse of $\mathcal{F}=13,000$ and linewidth (HWHM) of $2\pi\times500$ kHz.

A 1064 nm Nd:YAG laser beam is passed through an amplitude modulator (AM), a phase modulator (PM), and a second AM before it is injected into the optomechanical cavity. The cavity length is manually controlled by tuning the voltage sent to the piezoelectric device (PZT) on the macroscopic output coupler until the cavity is near resonance. 
Initially, the cavity is locked with the transmitted light that is incident on the locking diode,  $\mathrm{PD_L}$.
The signal from $\mathrm{PD_L}$ is conditioned by a servo amplifier (SA) before being sent to the second AM to lock the cavity. The optomechancial cavity is locked at a detuning of about 0.6 linewidths to stabilize the cavity using a strong optical spring with a resonance frequency above 100 kHz \cite{Cripe_RPL}. 
The light reflected from the cavity is detected by the measurement photodetector, $\mathrm{PD_M}$. The signal from $\mathrm{PD_M}$ is passed through a separate SA before being sent to the PM. After the cavity is locked with both the AM and PM feedback loops, the gain on the SA in the $\mathrm{PD_L}$ feedback loop is turned down to zero. The cavity remains locked using the PM feedback loop. The final measurement configuration uses only the reflected light because the transmitted light has relatively large shot noise due to the small transmission ($\approx 50$ ppm) of the end mirror, which may pollute the measurement. The reflection locking with the PM is less robust, and we are not able to directly acquire lock without first using the transmission locking. A separate intensity stabilization servo (ISS) loop and secondary servo are used to stabilize the power incident on the cavity to shot noise. 



Thermal noise and quantum noise must be modeled to fully account for the measured noise in the experiment. Thermal noise, which is governed by the fluctuation dissipation theorem \cite{FDT_Nyquist}, sets a limit on the precision of mechanical experiments \cite{FDT_Callen_1}. Thermal noise models are loosely divided into viscous or velocity-dependent models and internal friction models, depending on the source of dissipation. 
Structural damping models contain a frequency independent loss angle, $\phi$, and for a harmonic oscillator have a displacement amplitude spectral density of
\begin{eqnarray}
x(\omega) = 
\sqrt{\frac{4\mathrm{k_B}T\omega_{\mathrm{m}}^2}{\omega m Q [(\omega_{\mathrm{m}}^2-\omega^2)^2+\frac{\omega_{\mathrm{m}}^4}{Q^2}]}} \label{TNS_equation}
\end{eqnarray}
where $\mathrm{k_B}$ is the Boltzmann constant, $T$ is temperature, $m$ is mass, $Q = \sfrac{1}{\phi}$ is the quality factor, $\omega = 2\pi\times f$, and $\omega_{\mathrm{m}}$ is the angular frequency of the mechanical mode \cite{Saulson}. For structural damping, the thermal noise falls off as $\sfrac{1}{\omega^{\sfrac{5}{2}}}$ above the mechanical resonance frequency. Viscous damping, on the other hand, is proportional to $\sfrac{1}{\omega^2}$ above the mechanical resonance \cite{Saulson}.

 

We model the thermal noise using a finite element model of the microresonator that is constrained by measurements of the frequencies and quality factors of the fundamental mode and the next three higher-order modes. We infer the modal mass for each mode by using the thermal noise measurement presented below and are able to reproduce the inferred modal masses using the finite element model.
The thermal noise spectrum is calculated using Eq. \ref{TNS_equation} by summing the contribution of each mechanical mode in quadrature.

\begin{figure*}
\center
\includegraphics[trim={0.5cm .1cm .2cm .2cm},clip,width=.9\textwidth]{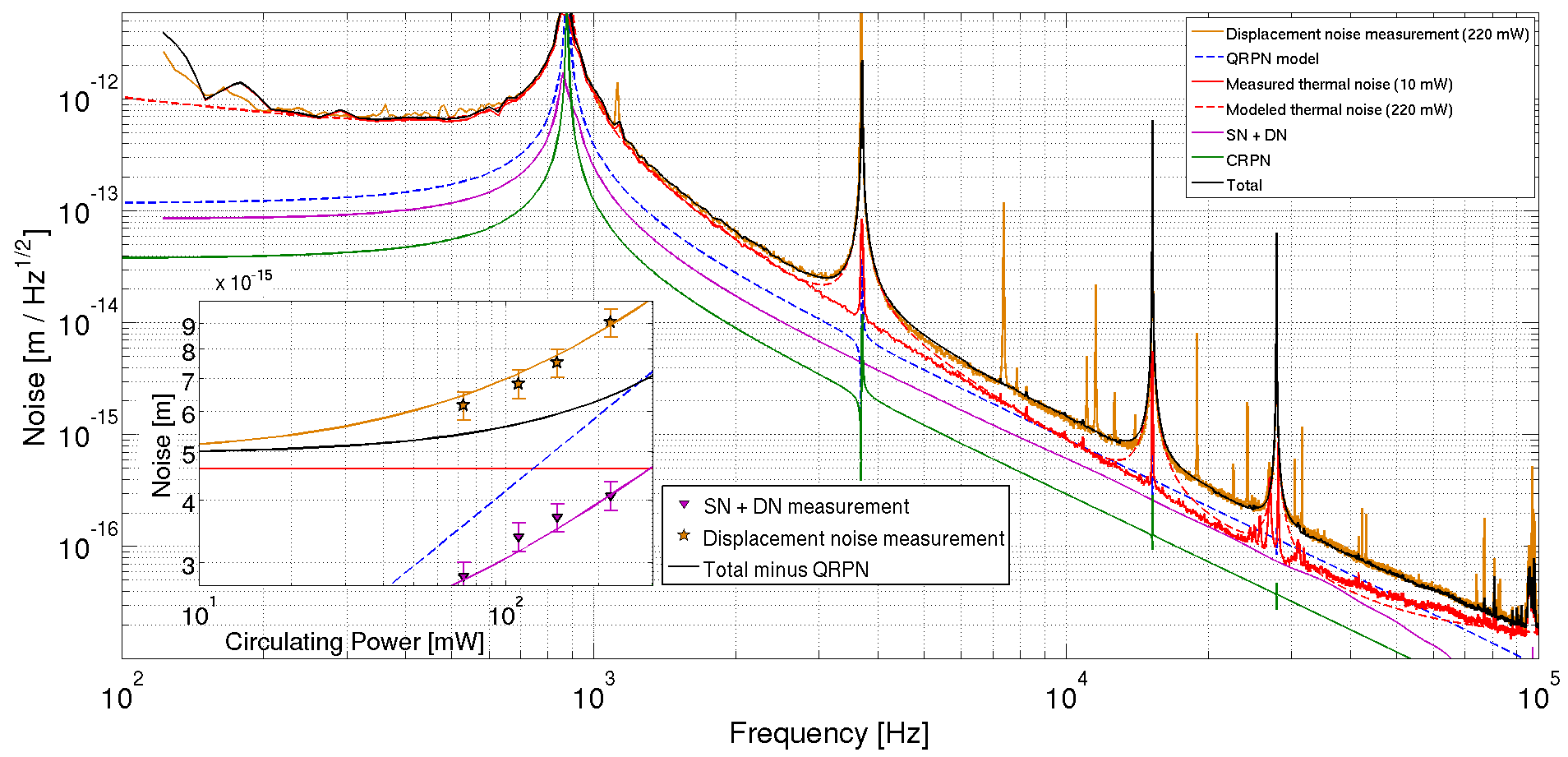}
\caption{Noise spectrum taken with a cavity circulating power of 220 mW as a function of frequency ($f$). The noise spectrum includes shot noise and dark noise, classical radiation pressure noise (CRPN), measured and modeled thermal noise, modeled QRPN, and a displacement noise measurement. The contributions from shot noise and dark noise, CRPN, thermal noise, and modeled QRPN are added in quadrature for a prediction of the total noise. The total noise includes a contribution from the measured thermal noise away from the resonances of the higher order mechanical modes and uses the modeled thermal noise around the resonance peaks to account for the change in mode coupling between low and high circulating power as described in the text. The thermal noise measurement was taken with 5\% of the circulating power used in the displacement noise measurement shown in orange. The peaks that are present in the displacement noise measurement that are not associated with a mechanical resonance are a result of nonlinear coupling that is not present at low circulating powers.
The inset includes four measurements at circulating powers of 73 mW, 110 mW, 150 mW, and 220 mW to show how each of the noise sources scale with cavity circulating power. Each type of noise shown in the noise spectrum is integrated over a 1 kHz frequency bin between 21 kHz and 22 kHz. The purple triangles are measurements of the shot noise plus dark noise, and the purple curve is a fit to that data. The stars correspond to the displacement noise measured for the four power levels and the red line is the measured thermal noise. The CRPN is not shown in the inset because it is well below the other noise sources. The orange curve in the inset is the expected total noise including the contributions from shot noise and dark noise, CRPN, thermal noise, and QRPN. The error bars on the measured data represent the measurement error based on the statistical uncertainty from multiple measurements.
The black curve is the expected total noise without including the contribution from QRPN.
}
\label{Spectrum_scaling}
\end{figure*}

The quantum noise is modeled using the input-output relations, which consist of a set of equations that relate the output fields to the input fields \cite{Corbitt_mathematical, 9}. 
Cavity losses are also included and serve as an input for vacuum fluctuations to enter the cavity. The dynamics of the microresonator are added based on the same finite element model that is used for the thermal noise model and measurements of the Q, frequency, and modal mass of each mechanical mode. This data is used to calculate the mechanical susceptibility of the microresonator. The cavity losses and detuning from resonance are determined by measurements of the optical spring. The input-output relations give the circulating power inside the cavity and the amount of light that is transmitted and reflected by the cavity for the carrier and sideband fields. The QRPN, shown in Figure \ref{Spectrum_scaling}, is calculated using these input-output relations and is proportional to $\sfrac{1}{\omega^{2}}$ above the mechanical resonance frequency.



After the cavity is locked, the signal from $\mathrm{PD_M}$ is sent to a spectrum analyzer for analysis. We measure an uncalibrated noise spectrum by first measuring the power spectral density of the error signal. We calibrate the spectrum by measuring the transfer function from the laser-cavity piezo to our error signal. The laser-cavity piezo changes the laser frequency and has been calibrated separately. We calibrate the error signal into length by measuring the response of the error signal to a change in the laser frequency and by knowing the cavity length.

To measure the thermal noise, we decrease the circulating power so that the QRPN is small compared to the Brownian motion of the microresonator.
The measurement of thermal noise, shown in Figure \ref{Spectrum_scaling} agrees with the structural damping model from 200 Hz to 30 kHz. 
We observe excess noise above 30 kHz, which may be related to thermoelastic damping, but is not entirely understood. The modeled thermal noise shown in Figure \ref{Spectrum_scaling} includes a thermoelastic damping contribution from the low Q drumhead mode that is between 5 MHz and 10 MHz based on the finite element model of the microresonator.
The measured noise also deviates from the model around the yaw mode at 3.7 kHz, pitch mode at 15 kHz, and side-to-side mode at 28 kHz as the coupling of these modes changes as a function of power due to radiation pressure induced torque. The coupling to all three modes was minimized for a low circulating power when the thermal noise measurement was taken. At higher circulating powers the radiation pressure force from the light is enough to bend the cantilever structure and cause the beam to hit the mirror at a slightly different location, enhancing the optomechanical coupling.

We take measurements at four cavity circulating powers of 73 mW, 110 mW, 150 mW, and 220 mW while maintaining a constant cavity detuning of about $0.6$ linewidths. The detuning of 0.6 linewidths is chosen to maximize the circulating power inside the cavity while maintaining a stable feedback loop to lock the cavity. The optical spring frequencies for the four measurements are 104 kHz, 119 kHz, 131 kHz, and 137 kHz, respectively. The noise spectrum of the 220 mW measurement is shown in Figure \ref{Spectrum_scaling}. 
The noise spectrum shows that QRPN is the largest displacement noise source between 10 kHz and 50 kHz. Below 10 kHz, thermal noise is the biggest contributor to the displacement noise, but the effect of QRPN is still visible in the displacement noise measurement down to 2 kHz, where it accounts for about 20\% of the measured displacement noise.

The data from each of the four displacement noise measurements, along with other known noise sources, is shown in the inset in Figure \ref{Spectrum_scaling} where it has been integrated over a 1 kHz band between 21 kHz and 22 kHz. 
QRPN scales with the square root of power \cite{Caves_1980} and is the largest noise source for circulating powers above 150 mW. For the measurement at 220 mW shown in the inset of Figure \ref{Spectrum_scaling}, QRPN represents 47\% of the total noise, while the thermal noise and shot noise plus dark noise contribute 30\% and 23\% of the total noise, respectively. Thermal noise is the biggest noise source for circulating powers below 150 mW, but again, QRPN still contributes to the total displacement noise at circulating powers of 10 mW and below. For each of the measurements shown in the inset of Figure \ref{Spectrum_scaling}, the sum of shot noise and dark noise is the third largest noise source.
While shot noise scales with the square root of power, the purple triangles shown in the inset of Figure \ref{Spectrum_scaling} include a contribution from dark noise and thus deviate from the expected scaling. While it may be counter intuitive that shot noise (calibrated to length) increases with power, this scaling is well understood as a result of the optical spring suppressing the signal \cite{9}. 
We sum the contribution of each of the noise sources to compute the total expected noise.
We find that our four displacement noise measurements, shown as orange stars in the inset, agree with the total expected noise with the statistical measurement error taken into account. The measurement error is calculated by repeating the measurement multiple times and is dominated by the fluctuations in the transfer function measurement that is used to calibrate the spectrum.
The orange curve shown in the inset of Figure \ref{Spectrum_scaling} is the predicted displacement noise without a contribution from QRPN. All four measurements of the displacement noise rule out the model without QRPN.

One effect that might mimic QRPN is bulk heating of the cantilever mirror caused by the absorption of the laser light. Due to the structural damping observed in our device, the mirror motion is dominated by thermal noise below 10 kHz, while still being QRPN limited above. The low frequency thermal noise may be used as a thermometer to measure any heating as a result of higher power. To explain the factor of two increase in noise observed at 20 kHz between low and high power as a result of heating, the temperature would have had to increase by a factor of 4. We can rule out this large increase in temperature by observing that the  measured noise at frequencies dominated by thermal noise (between 1 kHz and 2 kHz for example) only increases by 2\%, which is within measurement uncertainty. 

In conclusion, we present a measurement of quantum radiation pressure noise in a broad band of frequencies far from resonance of the mechanical oscillator. The observed noise spectrum shows the motion of the micro-resonator is affected by QRPN between about 2 kHz to 90 kHz. Analyzing all known significant noise sources, we show that the QRPN is the largest contributor between 10 kHz and 50 kHz, and that it scales as the square root of the optical power, as expected for quantum noise.

Since the first proposals of interferometric GW detectors, QRPN has been known to present a fundamental limit to the low frequency sensitivity of GW detectors. For the past two decades, the measurement of QRPN at frequencies relevant for gravitational wave detectors has eluded increasingly sensitive experiments. Meanwhile, several proposals for reducing QRPN \cite{Braginsky547, Braginsky_speedmeter, Kimble, Harms, Filter, FD_SQZ, Glasgow_speedmeter} have been relegated to theoretical concepts without the means to experimentally test them. This measurement of QRPN at frequencies in the gravitational wave band opens up the possibility of experimental tests of QRPN-reduction schemes. In addition, we are currently within a factor of five of the standard quantum limit \cite{Braginsky_SQL} at room temperature, which paves the way for a sensitivity below the standard quantum limit by cryogenically cooling the microresonator to reduce the thermal noise.

Measurement back-action limits the sensitivity of all force and position measurement.  Moreover, QRPN buffets the mirrors of GW detectors, limiting the sensitivity to GWs. Measuring the QRPN over this broad frequency band enables experimental tests of QRPN-reduction schemes to manipulate and mitigate this vexing noise source. More fundamentally, this measurement amounts to observation of the quantum vacuum fluctuation moving a macroscopic object.

\section{Acknowledgments}
JC and TC are supported by the National Science Foundation grant PHY-1150531. NA, AL, and NM are supported by National Science Foundation grants PHY-1707840 and PHY-1404245. A portion of this work was performed in the UCSB Nanofabrication Facility. JC and TC would like to thank David McClelland, Robert Ward, and Min Jet Yap for thoughtful comments and discussion. The authors would like to thank Christopher Wipf and Vivishek Sudhir for comments during the manuscript preparation. This document has been assigned the LIGO document number LIGO-P1800033. 

\section{Author contributions}
JC led the design, construction, and data taking for the experiment, which is part of his doctoral thesis, and prepared the manuscript. TC supervised the design and construction of the experiment and also helped with the data taking and analysis of the results. NA, RL, AL, and NM contributed to the design of the microresonators. GDC provided feedback on the design of the microresonators and, together with PH and DF, fabricated the devices. RS contributed experience from previous experiments. 


\begin{thebibliography}{10}%
\makeatletter
\providecommand \@ifxundefined [1]{%
 \@ifx{#1\undefined}
}%
\providecommand \@ifnum [1]{%
 \ifnum #1\expandafter \@firstoftwo
 \else \expandafter \@secondoftwo
 \fi
}%
\providecommand \@ifx [1]{%
 \ifx #1\expandafter \@firstoftwo
 \else \expandafter \@secondoftwo
 \fi
}%
\providecommand \natexlab [1]{#1}%
\providecommand \enquote  [1]{``#1''}%
\providecommand \bibnamefont  [1]{#1}%
\providecommand \bibfnamefont [1]{#1}%
\providecommand \citenamefont [1]{#1}%
\providecommand \href@noop [0]{\@secondoftwo}%
\providecommand \href [0]{\begingroup \@sanitize@url \@href}%
\providecommand \@href[1]{\@@startlink{#1}\@@href}%
\providecommand \@@href[1]{\endgroup#1\@@endlink}%
\providecommand \@sanitize@url [0]{\catcode `\\12\catcode `\$12\catcode
  `\&12\catcode `\#12\catcode `\^12\catcode `\_12\catcode `\%12\relax}%
\providecommand \@@startlink[1]{}%
\providecommand \@@endlink[0]{}%
\providecommand \url  [0]{\begingroup\@sanitize@url \@url }%
\providecommand \@url [1]{\endgroup\@href {#1}{\urlprefix }}%
\providecommand \urlprefix  [0]{URL }%
\providecommand \Eprint [0]{\href }%
\providecommand \doibase [0]{http://dx.doi.org/}%
\providecommand \selectlanguage [0]{\@gobble}%
\providecommand \bibinfo  [0]{\@secondoftwo}%
\providecommand \bibfield  [0]{\@secondoftwo}%
\providecommand \translation [1]{[#1]}%
\providecommand \BibitemOpen [0]{}%
\providecommand \bibitemStop [0]{}%
\providecommand \bibitemNoStop [0]{.\EOS\space}%
\providecommand \EOS [0]{\spacefactor3000\relax}%
\providecommand \BibitemShut  [1]{\csname bibitem#1\endcsname}%
\let\auto@bib@innerbib\@empty
\bibitem [{\citenamefont {Caves}(1980)}]{Caves_1980}%
  \BibitemOpen
  \bibfield  {author} {\bibinfo {author} {\bibfnamefont {C.~M.}\ \bibnamefont
  {Caves}},\ }\href {\doibase 10.1103/PhysRevLett.45.75} {\bibfield  {journal}
  {\bibinfo  {journal} {Phys. Rev. Lett.}\ }\textbf {\bibinfo {volume} {45}},\
  \bibinfo {pages} {75} (\bibinfo {year} {1980})}\BibitemShut {NoStop}%
\bibitem [{\citenamefont {{(LIGO Scientific Collaboration)}}(2015)}]{LIGO}%
  \BibitemOpen
  \bibfield  {author} {\bibinfo {author} {\bibfnamefont {{\relax J. Aasi
  \textit{et al}}.}~\bibnamefont {{(LIGO Scientific Collaboration)}}},\ }\href
  {http://stacks.iop.org/0264-9381/32/i=7/a=074001} {\bibfield  {journal}
  {\bibinfo  {journal} {Classical and Quantum Gravity}\ }\textbf {\bibinfo
  {volume} {32}},\ \bibinfo {pages} {074001} (\bibinfo {year}
  {2015})}\BibitemShut {NoStop}%
\bibitem [{\citenamefont {{(VIRGO Collaboration)}}(2015)}]{VIRGO}%
  \BibitemOpen
  \bibfield  {author} {\bibinfo {author} {\bibfnamefont {{\relax F. Acernese
  \textit{et al}}.}~\bibnamefont {{(VIRGO Collaboration)}}},\ }\href
  {http://stacks.iop.org/0264-9381/32/i=2/a=024001} {\bibfield  {journal}
  {\bibinfo  {journal} {Classical and Quantum Gravity}\ }\textbf {\bibinfo
  {volume} {32}},\ \bibinfo {pages} {024001} (\bibinfo {year}
  {2015})}\BibitemShut {NoStop}%
\bibitem [{\citenamefont {{(KAGRA Collaboration)}}(2012)}]{KAGRA}%
  \BibitemOpen
  \bibfield  {author} {\bibinfo {author} {\bibfnamefont {{\relax K. Somiya
  \textit{et al}}.}~\bibnamefont {{(KAGRA Collaboration)}}},\ }\href
  {http://stacks.iop.org/0264-9381/29/i=12/a=124007} {\bibfield  {journal}
  {\bibinfo  {journal} {Classical and Quantum Gravity}\ }\textbf {\bibinfo
  {volume} {29}},\ \bibinfo {pages} {124007} (\bibinfo {year}
  {2012})}\BibitemShut {NoStop}%
\bibitem [{\citenamefont {Braginsky}\ \emph {et~al.}(1980)\citenamefont
  {Braginsky}, \citenamefont {Vorontsov},\ and\ \citenamefont
  {Thorne}}]{Braginsky547}%
  \BibitemOpen
  \bibfield  {author} {\bibinfo {author} {\bibfnamefont {V.~B.}\ \bibnamefont
  {Braginsky}}, \bibinfo {author} {\bibfnamefont {Y.~I.}\ \bibnamefont
  {Vorontsov}}, \ and\ \bibinfo {author} {\bibfnamefont {K.~S.}\ \bibnamefont
  {Thorne}},\ }\href {\doibase 10.1126/science.209.4456.547} {\bibfield
  {journal} {\bibinfo  {journal} {Science}\ }\textbf {\bibinfo {volume}
  {209}},\ \bibinfo {pages} {547} (\bibinfo {year} {1980})}\BibitemShut
  {NoStop}%
\bibitem [{\citenamefont {Braginsky}\ \emph {et~al.}(2000)\citenamefont
  {Braginsky}, \citenamefont {Gorodetsky}, \citenamefont {Khalili},\ and\
  \citenamefont {Thorne}}]{Braginsky_speedmeter}%
  \BibitemOpen
  \bibfield  {author} {\bibinfo {author} {\bibfnamefont {V.~B.}\ \bibnamefont
  {Braginsky}}, \bibinfo {author} {\bibfnamefont {M.~L.}\ \bibnamefont
  {Gorodetsky}}, \bibinfo {author} {\bibfnamefont {F.~Y.}\ \bibnamefont
  {Khalili}}, \ and\ \bibinfo {author} {\bibfnamefont {K.~S.}\ \bibnamefont
  {Thorne}},\ }\href {\doibase 10.1103/PhysRevD.61.044002} {\bibfield
  {journal} {\bibinfo  {journal} {Phys. Rev. D}\ }\textbf {\bibinfo {volume}
  {61}},\ \bibinfo {pages} {044002} (\bibinfo {year} {2000})}\BibitemShut
  {NoStop}%
\bibitem [{\citenamefont {Kimble}\ \emph {et~al.}(2001)\citenamefont {Kimble},
  \citenamefont {Levin}, \citenamefont {Matsko}, \citenamefont {Thorne},\ and\
  \citenamefont {Vyatchanin}}]{Kimble}%
  \BibitemOpen
  \bibfield  {author} {\bibinfo {author} {\bibfnamefont {H.~J.}\ \bibnamefont
  {Kimble}}, \bibinfo {author} {\bibfnamefont {Y.}~\bibnamefont {Levin}},
  \bibinfo {author} {\bibfnamefont {A.~B.}\ \bibnamefont {Matsko}}, \bibinfo
  {author} {\bibfnamefont {K.~S.}\ \bibnamefont {Thorne}}, \ and\ \bibinfo
  {author} {\bibfnamefont {S.~P.}\ \bibnamefont {Vyatchanin}},\ }\href
  {\doibase 10.1103/PhysRevD.65.022002} {\bibfield  {journal} {\bibinfo
  {journal} {Phys. Rev. D}\ }\textbf {\bibinfo {volume} {65}},\ \bibinfo
  {pages} {022002} (\bibinfo {year} {2001})}\BibitemShut {NoStop}%
\bibitem [{\citenamefont {Harms}\ \emph {et~al.}(2003)\citenamefont {Harms},
  \citenamefont {Chen}, \citenamefont {Chelkowski}, \citenamefont {Franzen},
  \citenamefont {Vahlbruch}, \citenamefont {Danzmann},\ and\ \citenamefont
  {Schnabel}}]{Harms}%
  \BibitemOpen
  \bibfield  {author} {\bibinfo {author} {\bibfnamefont {J.}~\bibnamefont
  {Harms}}, \bibinfo {author} {\bibfnamefont {Y.}~\bibnamefont {Chen}},
  \bibinfo {author} {\bibfnamefont {S.}~\bibnamefont {Chelkowski}}, \bibinfo
  {author} {\bibfnamefont {A.}~\bibnamefont {Franzen}}, \bibinfo {author}
  {\bibfnamefont {H.}~\bibnamefont {Vahlbruch}}, \bibinfo {author}
  {\bibfnamefont {K.}~\bibnamefont {Danzmann}}, \ and\ \bibinfo {author}
  {\bibfnamefont {R.}~\bibnamefont {Schnabel}},\ }\href {\doibase
  10.1103/PhysRevD.68.042001} {\bibfield  {journal} {\bibinfo  {journal} {Phys.
  Rev. D}\ }\textbf {\bibinfo {volume} {68}},\ \bibinfo {pages} {042001}
  (\bibinfo {year} {2003})}\BibitemShut {NoStop}%
\bibitem [{\citenamefont {Evans}\ \emph {et~al.}(2013)\citenamefont {Evans},
  \citenamefont {Barsotti}, \citenamefont {Kwee}, \citenamefont {Harms},\ and\
  \citenamefont {Miao}}]{Filter}%
  \BibitemOpen
  \bibfield  {author} {\bibinfo {author} {\bibfnamefont {M.}~\bibnamefont
  {Evans}}, \bibinfo {author} {\bibfnamefont {L.}~\bibnamefont {Barsotti}},
  \bibinfo {author} {\bibfnamefont {P.}~\bibnamefont {Kwee}}, \bibinfo {author}
  {\bibfnamefont {J.}~\bibnamefont {Harms}}, \ and\ \bibinfo {author}
  {\bibfnamefont {H.}~\bibnamefont {Miao}},\ }\href {\doibase
  10.1103/PhysRevD.88.022002} {\bibfield  {journal} {\bibinfo  {journal} {Phys.
  Rev. D}\ }\textbf {\bibinfo {volume} {88}},\ \bibinfo {pages} {022002}
  (\bibinfo {year} {2013})}\BibitemShut {NoStop}%
\bibitem [{\citenamefont {Oelker}\ \emph {et~al.}(2016)\citenamefont {Oelker},
  \citenamefont {Isogai}, \citenamefont {Miller}, \citenamefont {Tse},
  \citenamefont {Barsotti}, \citenamefont {Mavalvala},\ and\ \citenamefont
  {Evans}}]{FD_SQZ}%
  \BibitemOpen
  \bibfield  {author} {\bibinfo {author} {\bibfnamefont {E.}~\bibnamefont
  {Oelker}}, \bibinfo {author} {\bibfnamefont {T.}~\bibnamefont {Isogai}},
  \bibinfo {author} {\bibfnamefont {J.}~\bibnamefont {Miller}}, \bibinfo
  {author} {\bibfnamefont {M.}~\bibnamefont {Tse}}, \bibinfo {author}
  {\bibfnamefont {L.}~\bibnamefont {Barsotti}}, \bibinfo {author}
  {\bibfnamefont {N.}~\bibnamefont {Mavalvala}}, \ and\ \bibinfo {author}
  {\bibfnamefont {M.}~\bibnamefont {Evans}},\ }\href {\doibase
  10.1103/PhysRevLett.116.041102} {\bibfield  {journal} {\bibinfo  {journal}
  {Phys. Rev. Lett.}\ }\textbf {\bibinfo {volume} {116}},\ \bibinfo {pages}
  {041102} (\bibinfo {year} {2016})}\BibitemShut {NoStop}%
\bibitem [{\citenamefont {\relax C.~Gräf~\textit{et
  al}.}(2014)}]{Glasgow_speedmeter}%
  \BibitemOpen
  \bibfield  {author} {\bibinfo {author} {\bibnamefont {\relax
  C.~Gräf~\textit{et al}.}},\ }\href {\doibase
  https://doi.org/10.1088/0264-9381/31/21/215009} {\bibfield  {journal}
  {\bibinfo  {journal} {Class. Quantum Grav.}\ }\textbf {\bibinfo {volume}
  {31}},\ \bibinfo {pages} {215009} (\bibinfo {year} {2014})}\BibitemShut
  {NoStop}%
\bibitem [{\citenamefont {Braginsky}\ and\ \citenamefont
  {Manukin}(1977)}]{Braginsky_book}%
  \BibitemOpen
  \bibfield  {author} {\bibinfo {author} {\bibfnamefont {V.~B.}\ \bibnamefont
  {Braginsky}}\ and\ \bibinfo {author} {\bibfnamefont {A.~B.}\ \bibnamefont
  {Manukin}},\ }\href@noop {} {\emph {\bibinfo {title} {Measurement of Weak
  Forces in Physics Experiments}}}\ (\bibinfo  {publisher} {University of
  Chicago Press},\ \bibinfo {year} {1977})\BibitemShut {NoStop}%
\bibitem [{\citenamefont {Caves}(1981)}]{Caves1981}%
  \BibitemOpen
  \bibfield  {author} {\bibinfo {author} {\bibfnamefont {C.~M.}\ \bibnamefont
  {Caves}},\ }\href {\doibase 10.1103/PhysRevD.23.1693} {\bibfield  {journal}
  {\bibinfo  {journal} {Phys. Rev. D}\ }\textbf {\bibinfo {volume} {23}},\
  \bibinfo {pages} {1693} (\bibinfo {year} {1981})}\BibitemShut {NoStop}%
\bibitem [{\citenamefont {Saulson}(1990)}]{Saulson}%
  \BibitemOpen
  \bibfield  {author} {\bibinfo {author} {\bibfnamefont {P.~R.}\ \bibnamefont
  {Saulson}},\ }\href {\doibase 10.1103/PhysRevD.42.2437} {\bibfield  {journal}
  {\bibinfo  {journal} {Phys. Rev. D}\ }\textbf {\bibinfo {volume} {42}},\
  \bibinfo {pages} {2437} (\bibinfo {year} {1990})}\BibitemShut {NoStop}%
\bibitem [{\citenamefont {Purdy}\ \emph {et~al.}(2013)\citenamefont {Purdy},
  \citenamefont {Peterson},\ and\ \citenamefont {Regal}}]{Purdy}%
  \BibitemOpen
  \bibfield  {author} {\bibinfo {author} {\bibfnamefont {T.~P.}\ \bibnamefont
  {Purdy}}, \bibinfo {author} {\bibfnamefont {R.~W.}\ \bibnamefont {Peterson}},
  \ and\ \bibinfo {author} {\bibfnamefont {C.~A.}\ \bibnamefont {Regal}},\
  }\href {\doibase 10.1126/science.1231282} {\bibfield  {journal} {\bibinfo
  {journal} {Science}\ }\textbf {\bibinfo {volume} {339}},\ \bibinfo {pages}
  {801} (\bibinfo {year} {2013})}\BibitemShut {NoStop}%
\bibitem [{\citenamefont {Teufel}\ \emph {et~al.}(2016)\citenamefont {Teufel},
  \citenamefont {Lecocq},\ and\ \citenamefont {Simmonds}}]{Teufel}%
  \BibitemOpen
  \bibfield  {author} {\bibinfo {author} {\bibfnamefont {J.~D.}\ \bibnamefont
  {Teufel}}, \bibinfo {author} {\bibfnamefont {F.}~\bibnamefont {Lecocq}}, \
  and\ \bibinfo {author} {\bibfnamefont {R.~W.}\ \bibnamefont {Simmonds}},\
  }\href {\doibase 10.1103/PhysRevLett.116.013602} {\bibfield  {journal}
  {\bibinfo  {journal} {Phys. Rev. Lett.}\ }\textbf {\bibinfo {volume} {116}},\
  \bibinfo {pages} {013602} (\bibinfo {year} {2016})}\BibitemShut {NoStop}%
\bibitem [{\citenamefont {Purdy}\ \emph {et~al.}(2017)\citenamefont {Purdy},
  \citenamefont {Grutter}, \citenamefont {Srinivasan},\ and\ \citenamefont
  {Taylor}}]{Purdy_room_T}%
  \BibitemOpen
  \bibfield  {author} {\bibinfo {author} {\bibfnamefont {T.~P.}\ \bibnamefont
  {Purdy}}, \bibinfo {author} {\bibfnamefont {K.~E.}\ \bibnamefont {Grutter}},
  \bibinfo {author} {\bibfnamefont {K.}~\bibnamefont {Srinivasan}}, \ and\
  \bibinfo {author} {\bibfnamefont {J.~M.}\ \bibnamefont {Taylor}},\ }\href
  {\doibase 10.1126/science.aag1407} {\bibfield  {journal} {\bibinfo  {journal}
  {Science}\ }\textbf {\bibinfo {volume} {356}},\ \bibinfo {pages} {1265}
  (\bibinfo {year} {2017})}\BibitemShut {NoStop}%
\bibitem [{\citenamefont {Sudhir}\ \emph {et~al.}(2017)\citenamefont {Sudhir},
  \citenamefont {Schilling}, \citenamefont {Fedorov}, \citenamefont {Sch\"utz},
  \citenamefont {Wilson},\ and\ \citenamefont {Kippenberg}}]{Sudhir}%
  \BibitemOpen
  \bibfield  {author} {\bibinfo {author} {\bibfnamefont {V.}~\bibnamefont
  {Sudhir}}, \bibinfo {author} {\bibfnamefont {R.}~\bibnamefont {Schilling}},
  \bibinfo {author} {\bibfnamefont {S.~A.}\ \bibnamefont {Fedorov}}, \bibinfo
  {author} {\bibfnamefont {H.}~\bibnamefont {Sch\"utz}}, \bibinfo {author}
  {\bibfnamefont {D.~J.}\ \bibnamefont {Wilson}}, \ and\ \bibinfo {author}
  {\bibfnamefont {T.~J.}\ \bibnamefont {Kippenberg}},\ }\href {\doibase
  10.1103/PhysRevX.7.031055} {\bibfield  {journal} {\bibinfo  {journal} {Phys.
  Rev. X}\ }\textbf {\bibinfo {volume} {7}},\ \bibinfo {pages} {031055}
  (\bibinfo {year} {2017})}\BibitemShut {NoStop}%
\bibitem [{\citenamefont {Cole}\ \emph {et~al.}(2008)\citenamefont {Cole},
  \citenamefont {Gr{\"o}blacher}, \citenamefont {Gugler}, \citenamefont
  {Gigan},\ and\ \citenamefont {Aspelmeyer}}]{cole08}%
  \BibitemOpen
  \bibfield  {author} {\bibinfo {author} {\bibfnamefont {G.~D.}\ \bibnamefont
  {Cole}}, \bibinfo {author} {\bibfnamefont {S.}~\bibnamefont
  {Gr{\"o}blacher}}, \bibinfo {author} {\bibfnamefont {K.}~\bibnamefont
  {Gugler}}, \bibinfo {author} {\bibfnamefont {S.}~\bibnamefont {Gigan}}, \
  and\ \bibinfo {author} {\bibfnamefont {M.}~\bibnamefont {Aspelmeyer}},\
  }\href {\doibase 10.1063/1.2952512} {\bibfield  {journal} {\bibinfo
  {journal} {Applied Physics Letters}\ }\textbf {\bibinfo {volume} {92}},\
  \bibinfo {eid} {261108} (\bibinfo {year} {2008})}\BibitemShut {NoStop}%
\bibitem [{\citenamefont {Cole}(2012)}]{cole12}%
  \BibitemOpen
  \bibfield  {author} {\bibinfo {author} {\bibfnamefont {G.~D.}\ \bibnamefont
  {Cole}},\ }in\ \href {\doibase 10.1117/12.931226} {\emph {\bibinfo
  {booktitle} {Proc. SPIE 8458, Optics \& Photonics, Optical Trapping and
  Optical Micromanipulation IX}}}\ (\bibinfo  {publisher} {SPIE},\ \bibinfo
  {year} {2012})\ p.\ \bibinfo {pages} {845807}\BibitemShut {NoStop}%
\bibitem [{\citenamefont {Cole}\ \emph {et~al.}(2013)\citenamefont {Cole},
  \citenamefont {Zhang}, \citenamefont {Martin}, \citenamefont {Ye},\ and\
  \citenamefont {Aspelmeyer}}]{cole13}%
  \BibitemOpen
  \bibfield  {author} {\bibinfo {author} {\bibfnamefont {G.~D.}\ \bibnamefont
  {Cole}}, \bibinfo {author} {\bibfnamefont {W.}~\bibnamefont {Zhang}},
  \bibinfo {author} {\bibfnamefont {M.~J.}\ \bibnamefont {Martin}}, \bibinfo
  {author} {\bibfnamefont {J.}~\bibnamefont {Ye}}, \ and\ \bibinfo {author}
  {\bibfnamefont {M.}~\bibnamefont {Aspelmeyer}},\ }\href
  {http://dx.doi.org/10.1038/nphoton.2013.174} {\bibfield  {journal} {\bibinfo
  {journal} {Nat Photon}\ }\textbf {\bibinfo {volume} {7}},\ \bibinfo {pages}
  {644} (\bibinfo {year} {2013})}\BibitemShut {NoStop}%
\bibitem [{\citenamefont {Cole}\ \emph {et~al.}(2016)\citenamefont {Cole},
  \citenamefont {Zhang}, \citenamefont {Bjork}, \citenamefont {Follman},
  \citenamefont {Heu}, \citenamefont {Deutsch}, \citenamefont {Sonderhouse},
  \citenamefont {Robinson}, \citenamefont {Franz}, \citenamefont
  {Alexandrovski}, \citenamefont {Notcutt}, \citenamefont {Heckl},
  \citenamefont {Ye},\ and\ \citenamefont {Aspelmeyer}}]{cole14}%
  \BibitemOpen
  \bibfield  {author} {\bibinfo {author} {\bibfnamefont {G.~D.}\ \bibnamefont
  {Cole}}, \bibinfo {author} {\bibfnamefont {W.}~\bibnamefont {Zhang}},
  \bibinfo {author} {\bibfnamefont {B.~J.}\ \bibnamefont {Bjork}}, \bibinfo
  {author} {\bibfnamefont {D.}~\bibnamefont {Follman}}, \bibinfo {author}
  {\bibfnamefont {P.}~\bibnamefont {Heu}}, \bibinfo {author} {\bibfnamefont
  {C.}~\bibnamefont {Deutsch}}, \bibinfo {author} {\bibfnamefont
  {L.}~\bibnamefont {Sonderhouse}}, \bibinfo {author} {\bibfnamefont
  {J.}~\bibnamefont {Robinson}}, \bibinfo {author} {\bibfnamefont
  {C.}~\bibnamefont {Franz}}, \bibinfo {author} {\bibfnamefont
  {A.}~\bibnamefont {Alexandrovski}}, \bibinfo {author} {\bibfnamefont
  {M.}~\bibnamefont {Notcutt}}, \bibinfo {author} {\bibfnamefont {O.~H.}\
  \bibnamefont {Heckl}}, \bibinfo {author} {\bibfnamefont {J.}~\bibnamefont
  {Ye}}, \ and\ \bibinfo {author} {\bibfnamefont {M.}~\bibnamefont
  {Aspelmeyer}},\ }\href {\doibase 10.1364/OPTICA.3.000647} {\bibfield
  {journal} {\bibinfo  {journal} {Optica}\ }\textbf {\bibinfo {volume} {3}},\
  \bibinfo {pages} {647} (\bibinfo {year} {2016})}\BibitemShut {NoStop}%
\bibitem [{\citenamefont {Singh}\ \emph {et~al.}(2016)\citenamefont {Singh},
  \citenamefont {Cole}, \citenamefont {Cripe},\ and\ \citenamefont
  {Corbitt}}]{Singh_PRL}%
  \BibitemOpen
  \bibfield  {author} {\bibinfo {author} {\bibfnamefont {R.}~\bibnamefont
  {Singh}}, \bibinfo {author} {\bibfnamefont {G.~D.}\ \bibnamefont {Cole}},
  \bibinfo {author} {\bibfnamefont {J.}~\bibnamefont {Cripe}}, \ and\ \bibinfo
  {author} {\bibfnamefont {T.}~\bibnamefont {Corbitt}},\ }\href {\doibase
  10.1103/PhysRevLett.117.213604} {\bibfield  {journal} {\bibinfo  {journal}
  {Phys. Rev. Lett.}\ }\textbf {\bibinfo {volume} {117}},\ \bibinfo {pages}
  {213604} (\bibinfo {year} {2016})}\BibitemShut {NoStop}%
\bibitem [{\citenamefont {Cripe}\ \emph {et~al.}(2018)\citenamefont {Cripe},
  \citenamefont {Aggarwal}, \citenamefont {Singh}, \citenamefont {Lanza},
  \citenamefont {Libson}, \citenamefont {Yap}, \citenamefont {Cole},
  \citenamefont {McClelland}, \citenamefont {Mavalvala},\ and\ \citenamefont
  {Corbitt}}]{Cripe_RPL}%
  \BibitemOpen
  \bibfield  {author} {\bibinfo {author} {\bibfnamefont {J.}~\bibnamefont
  {Cripe}}, \bibinfo {author} {\bibfnamefont {N.}~\bibnamefont {Aggarwal}},
  \bibinfo {author} {\bibfnamefont {R.}~\bibnamefont {Singh}}, \bibinfo
  {author} {\bibfnamefont {R.}~\bibnamefont {Lanza}}, \bibinfo {author}
  {\bibfnamefont {A.}~\bibnamefont {Libson}}, \bibinfo {author} {\bibfnamefont
  {M.~J.}\ \bibnamefont {Yap}}, \bibinfo {author} {\bibfnamefont {G.~D.}\
  \bibnamefont {Cole}}, \bibinfo {author} {\bibfnamefont {D.~E.}\ \bibnamefont
  {McClelland}}, \bibinfo {author} {\bibfnamefont {N.}~\bibnamefont
  {Mavalvala}}, \ and\ \bibinfo {author} {\bibfnamefont {T.}~\bibnamefont
  {Corbitt}},\ }\href {\doibase 10.1103/PhysRevA.97.013827} {\bibfield
  {journal} {\bibinfo  {journal} {Phys. Rev. A}\ }\textbf {\bibinfo {volume}
  {97}},\ \bibinfo {pages} {013827} (\bibinfo {year} {2018})}\BibitemShut
  {NoStop}%
\bibitem [{\citenamefont {Nyquist}(1928)}]{FDT_Nyquist}%
  \BibitemOpen
  \bibfield  {author} {\bibinfo {author} {\bibfnamefont {H.}~\bibnamefont
  {Nyquist}},\ }\href {\doibase 10.1103/PhysRev.32.110} {\bibfield  {journal}
  {\bibinfo  {journal} {Phys. Rev.}\ }\textbf {\bibinfo {volume} {32}},\
  \bibinfo {pages} {110} (\bibinfo {year} {1928})}\BibitemShut {NoStop}%
\bibitem [{\citenamefont {Callen}\ and\ \citenamefont
  {Welton}(1951)}]{FDT_Callen_1}%
  \BibitemOpen
  \bibfield  {author} {\bibinfo {author} {\bibfnamefont {H.~B.}\ \bibnamefont
  {Callen}}\ and\ \bibinfo {author} {\bibfnamefont {T.~A.}\ \bibnamefont
  {Welton}},\ }\href {\doibase 10.1103/PhysRev.83.34} {\bibfield  {journal}
  {\bibinfo  {journal} {Phys. Rev.}\ }\textbf {\bibinfo {volume} {83}},\
  \bibinfo {pages} {34} (\bibinfo {year} {1951})}\BibitemShut {NoStop}%
\bibitem [{\citenamefont {Corbitt}\ \emph {et~al.}(2005)\citenamefont
  {Corbitt}, \citenamefont {Chen},\ and\ \citenamefont
  {Mavalvala}}]{Corbitt_mathematical}%
  \BibitemOpen
  \bibfield  {author} {\bibinfo {author} {\bibfnamefont {T.}~\bibnamefont
  {Corbitt}}, \bibinfo {author} {\bibfnamefont {Y.}~\bibnamefont {Chen}}, \
  and\ \bibinfo {author} {\bibfnamefont {N.}~\bibnamefont {Mavalvala}},\ }\href
  {\doibase 10.1103/PhysRevA.72.013818} {\bibfield  {journal} {\bibinfo
  {journal} {Phys. Rev. A}\ }\textbf {\bibinfo {volume} {72}},\ \bibinfo
  {pages} {013818} (\bibinfo {year} {2005})}\BibitemShut {NoStop}%
\bibitem [{\citenamefont {Corbitt}\ \emph {et~al.}(2006)\citenamefont
  {Corbitt}, \citenamefont {Chen}, \citenamefont {Khalili}, \citenamefont
  {Ottaway}, \citenamefont {Vyatchanin}, \citenamefont {Whitcomb},\ and\
  \citenamefont {Mavalvala}}]{9}%
  \BibitemOpen
  \bibfield  {author} {\bibinfo {author} {\bibfnamefont {T.}~\bibnamefont
  {Corbitt}}, \bibinfo {author} {\bibfnamefont {Y.}~\bibnamefont {Chen}},
  \bibinfo {author} {\bibfnamefont {F.}~\bibnamefont {Khalili}}, \bibinfo
  {author} {\bibfnamefont {D.}~\bibnamefont {Ottaway}}, \bibinfo {author}
  {\bibfnamefont {S.}~\bibnamefont {Vyatchanin}}, \bibinfo {author}
  {\bibfnamefont {S.}~\bibnamefont {Whitcomb}}, \ and\ \bibinfo {author}
  {\bibfnamefont {N.}~\bibnamefont {Mavalvala}},\ }\href {\doibase
  10.1103/PhysRevA.73.023801} {\bibfield  {journal} {\bibinfo  {journal} {Phys.
  Rev. A}\ }\textbf {\bibinfo {volume} {73}},\ \bibinfo {pages} {023801}
  (\bibinfo {year} {2006})}\BibitemShut {NoStop}%
\bibitem [{\citenamefont {Braginsky}(1968)}]{Braginsky_SQL}%
  \BibitemOpen
  \bibfield  {author} {\bibinfo {author} {\bibfnamefont {V.~B.}\ \bibnamefont
  {Braginsky}},\ }\href@noop {} {\bibfield  {journal} {\bibinfo  {journal}
  {Sov. J. Exp. Theor. Phys.}\ }\textbf {\bibinfo {volume} {26}},\ \bibinfo
  {pages} {831} (\bibinfo {year} {1968})}\BibitemShut {NoStop}%
\end{thebibliography}

%

\end{document}